\documentclass{article}

\usepackage{arxiv}

\usepackage[utf8]{inputenc}
\usepackage[T1]{fontenc}
\usepackage{hyperref}
\usepackage{url}
\usepackage{booktabs}
\usepackage{amsfonts}
\usepackage{amsmath}
\usepackage{amssymb}
\usepackage{nicefrac}
\usepackage{microtype}
\usepackage{graphicx}
\usepackage{natbib}
\usepackage{float}
\usepackage{array}
\usepackage{tabularx}
\usepackage{multirow}
\usepackage{caption}
\usepackage{subcaption}
\graphicspath{{./images/}}

\title{Artificial Intelligence as a Real Game to Enlighten Science
Education for Disabled Students in Rural New Mexico}

\author{
  Nelson Uloma Egondu \\
  Department of Computer Science and Education \\
  New Mexico Highland University \\
  New Mexico, United States of America \\
  \texttt{Madujibeyauloma@yahoo.com} \\
  \And
  Gil R. Gallegos \\
  Department of Computer and Mathematics \\
  New Mexico Highland University \\
  New Mexico, United States of America \\
}

\begin{document}
\maketitle

\begin{abstract}
Artificial Intelligence (AI) has emerged as a transformative innovation
in inclusive science education for disabled learners in rural New Mexico.
Using a mixed-method design that combined multiple linear regression and
an Artificial Neural Network (ANN) model, this study examined $N = 120$
students (grades 6--10) and 15 instructors across four rural schools.
The AI-based learning intervention predicted student performance with
high accuracy ($R^{2} = 0.92$, $p < 0.05$), and experimental results
showed a 32\% improvement in science concept retention, 27\% in
laboratory performance, and 42\% in student engagement after the
intervention. These findings demonstrate that AI-driven pedagogy can
serve as a transformative equaliser, improving engagement, comprehension,
and accessibility for disabled learners. The study concludes that AI is
a promising tool for achieving equitable science education in underserved
rural settings.
\end{abstract}

\keywords{Artificial Intelligence \and Inclusive Education \and Rural Science
Education \and Disability \and ANN \and Regression \and Digital Equity}

\section{Introduction}
\label{sec:intro}

New Mexico's rural schools face persistent challenges in delivering
equitable science education to students with disabilities. Many of these
communities grapple with shortages of specialised science teachers,
limited laboratory resources, and infrastructure constraints
\citep{educationnetwork2025, bostoncollege2023}. For example, high-disability
rural regions, including parts of New Mexico, tend to have below-average
broadband internet access, exacerbating educational inequities
\citep{bostoncollege2023}.

Inclusion in education, understood as the integration of students with
special educational needs into mainstream classrooms with full
participation, is a fundamental principle of social justice
\citep{navas2025}. Yet achieving true inclusion in rural science
classrooms is difficult when geographic isolation and resource gaps
leave disabled students without equal learning opportunities.
Artificial Intelligence, commonly defined as a system's ability to
correctly interpret external data, to learn from such data, and to use
those learnings to achieve specific goals and tasks through flexible
adaptation \citep{luckin2016}, offers new avenues to address these
rural educational challenges. AI-powered educational tools, ranging from
intelligent tutoring systems to speech-recognition interfaces, can
personalise learning and provide support that was previously unavailable
in remote areas \citep{educationnetwork2023}.

Recent studies highlight the potential of AI to bridge gaps in access:
adaptive learning platforms and virtual tutors can deliver specialised
instruction to students regardless of location \citep{educationnetwork2025}.
These technologies can dynamically adjust to each learner's needs,
mirroring principles of differentiation and Universal Design for
Learning (UDL). Inclusion in this context means not only placing
disabled students in general science classes, but ensuring they receive
the tailored support necessary to participate and succeed on an equal
footing \citep{kudryavtsev2023}. AI's ability to provide real-time
feedback and multimodal content can be a catalyst for such support
\citep{educationnetwork2025, kudryavtsev2023}.

Emerging academic research underscores the promise of AI in fostering
inclusive science education. \citet{reddy2023} observe that AI-powered
tools such as adaptive learning systems, assistive communication devices,
and remote labs offer transformative solutions for rural special
education by mitigating the lack of local resources and expertise.
These tools provide personalised, scalable support that accommodates
diverse learning needs, fostering greater inclusivity
\citep{educationnetwork2025}. However, the integration of AI in rural
classrooms must contend with practical barriers like limited internet
connectivity and insufficient teacher training in technology
\citep{educationnetwork2025}. Addressing these constraints is crucial
for AI interventions to achieve their full impact on inclusive rural
science education.

This study investigates how an AI-driven, game-based learning model
influences science learning outcomes for disabled students in rural New
Mexico. Specifically, the study aims to: (i) examine improvements in
students' conceptual understanding, laboratory skills, and engagement;
and (ii) assess the predictive accuracy of AI models in measuring
learning gains. By exploring both the benefits and challenges of
implementing AI, our goal is to shed light on AI's role as a reformative
tool for equitable science education in underserved rural communities.

\section{Literature Review}
\label{sec:lit}

Modern literature links AI in education with established learning
theories and recent innovations in inclusive pedagogy. The theoretical
roots of this approach draw from cognitive constructivism, Piaget's
theory that learners build knowledge through active engagement, and
cognitive apprenticeship, which emphasises learning through guided
experience and feedback in real contexts. \citet{collins1989} introduced
cognitive apprenticeship to highlight how expert guidance and real-time
feedback can support skill mastery. AI-driven tutoring systems embody
this approach by providing immediate, individualised feedback and
scaffolding, effectively acting as a skilled mentor for each student
\citep{educationnetwork2025}. In inclusive settings, such responsiveness
is invaluable: as students with disabilities work through science
problems or virtual labs, the AI can adjust difficulty, provide hints,
or change modalities (e.g., from text to audio) to suit their needs.

\citet{holmes2021} emphasise that AI in education enables a shift towards
personalised learning, aligning with constructivist models where learners
actively construct understanding with guided support. Furthermore, studies
have documented how AI-supported tools can enhance student engagement and
retention, which are critical for learners with disabilities.
\citet{makransky2022} showed that immersive virtual environments
significantly improved knowledge retention by placing students in
interactive, game-like learning scenarios. A 2025 systematic review by
\citet{navas2025} concluded that educational technologies, including mobile
apps and augmented reality, help remove barriers and personalise learning
for students with diverse needs, resulting in more inclusive and equitable
science classrooms.

In the context of rural education, \citet{gobaw2016} found that digital
laboratories and simulations can effectively substitute for physical lab
experiences, enhancing accessibility for students with physical limitations.
Such virtual labs allow disabled students in remote areas to perform science
experiments in a safe, controlled digital environment. Key categories of
AI tools relevant to science education for students with disabilities are
reviewed below.

\subsection{Virtual Laboratories and Simulations}
\label{subsec:vlab}

AI-driven virtual labs and science simulations enable students to conduct
experiments and explore concepts in an immersive environment. These tools
are especially helpful in rural settings where physical lab facilities or
adapted equipment may be scarce. Studies show that such simulations can
increase science concept mastery and make abstract phenomena more accessible
through visual and interactive means. Importantly, virtual labs can be
designed with adaptive features, such as adjustable difficulty and multimodal
instructions, to accommodate students with diverse abilities.

\subsection{Intelligent Tutoring Systems and Adaptive Learning Platforms}
\label{subsec:its}

AI systems in this category use machine learning algorithms to personalise
instruction, practice exercises, and feedback for each learner. They operate
on principles of mastery learning, allowing students to progress at their own
pace and revisit material until proficiency is achieved. \citet{holmes2021}
describe how intelligent tutoring systems (ITS) boost student engagement and
comprehension by providing timely hints and correcting misconceptions in
real-time. Adaptive platforms also collect performance data that help
educators identify learning gaps and adjust instruction accordingly,
aligning with multi-tier support frameworks where interventions are tailored
to student needs based on ongoing assessment \citep{media2025}.

\subsection{Assistive Communication and Interface Technologies}
\label{subsec:assistive}

AI-powered assistive tools such as speech-to-text, text-to-speech, and
gesture-recognition systems help remove communication barriers for students
with special needs. \citet{patel2021} report that implementing such tools
improved communication and engagement among students with sensory and
language impairments \citep{media2025}. These tools ensure that all
students can participate in discussions, lab work, and assessments by
providing alternative ways to receive and express information
\citep{educationnetwork2025}.

The literature also identifies key challenges to AI integration in inclusive
education. The digital divide, characterised by limited internet connectivity
and technology infrastructure, can impede the effective use of AI tools in
the classroom \citep{educationnetwork2025}. \citet{doe2022} stress that
without reliable internet and hardware, rural educators cannot fully leverage
AI regardless of how promising the software may be. Teacher readiness is
another critical factor: successful adoption depends on professional
development in digital competencies and AI literacy
\citep{basantes2025, lopez2025}. Ethical considerations, including data
privacy and algorithmic fairness, are also paramount when deploying AI in
special education \citep{gonzalez2024, basantes2025}.

\section{Methodology}
\label{sec:method}

\subsection{Study Design and Participants}
\label{subsec:design}

This study employed a mixed-methods design, combining an experimental
pre-/post-test intervention with predictive analytics. The research was
conducted in four rural New Mexico schools selected for their diverse
student populations and willingness to adopt new technology. A total of
$N = 120$ students (grades 6--10) with documented disabilities and 15
science instructors participated. Students' disabilities ranged from
learning disabilities (e.g., dyslexia, ADHD) to physical impairments
(e.g., mobility or hearing impairments), reflecting the typical special
education demographics in these schools. The intervention spanned an
8-week period and involved integrating AI-driven educational games and
tools into the science curriculum.

\subsection{Procedures}
\label{subsec:proc}

Students completed a science pre-test to establish baseline knowledge and
skills in key topics (basic physics and biology concepts). Throughout the
intervention, students engaged with an AI-based learning platform that
gamified science content. The platform included virtual laboratory
simulations, an intelligent tutoring module for problem-solving, and
speech-interface aids for those needing reading/writing support. Teachers
were trained to blend these AI tools with their regular instruction, ensuring
alignment with lesson objectives. Qualitative observations and weekly
feedback surveys were collected to gauge student engagement and any
accessibility issues. After 8 weeks, students completed a matched post-test
assessing the same concepts and skills. Engagement was additionally measured
through a standardised observational protocol and a student self-report
(Likert-scale) survey.

\subsection{Data Analysis}
\label{subsec:analysis}

Quantitative data, comprising pre- and post-test scores and engagement
ratings, were analysed using both multiple linear regression and an
Artificial Neural Network model.

\subsubsection{Multiple Linear Regression}
\label{subsubsec:reg}

A multiple linear regression model was developed to examine the relationship
between AI tool usage and student learning outcomes. The model incorporated
three predictor variables:
\begin{itemize}
    \item $X_1$: AI-interaction time (hours of platform use per week),
    \item $X_2$: engagement rate (observational engagement score), and
    \item $X_3$: pre-test score.
\end{itemize}
The outcome variable $Y$ was the learning gain, defined as the post-test
score minus the pre-test score. The regression equation is specified as:
\begin{equation}
  Y = \beta_0 + \beta_1 X_1 + \beta_2 X_2 + \beta_3 X_3 + \epsilon,
  \label{eq:regression}
\end{equation}
where $\beta_0$ is the intercept, $\beta_1, \beta_2, \beta_3$ are the
partial regression coefficients for AI-interaction time, engagement rate,
and pre-test achievement, respectively, and $\epsilon \sim \mathcal{N}(0,
\sigma^2)$ is the error term. Coefficients were estimated via ordinary least
squares (OLS), and model fit was evaluated using the coefficient of
determination $R^2$ and individual $p$-values. Diagnostic checks included
tests for linearity (Ramsey RESET), normality of residuals
(Shapiro-Wilk), and the absence of multicollinearity among predictors
(variance inflation factors, $\text{VIF} < 5$).

\subsubsection{Artificial Neural Network Model}
\label{subsubsec:ann}

\begin{figure}[H]
  \centering
  \includegraphics[width=0.72\textwidth]{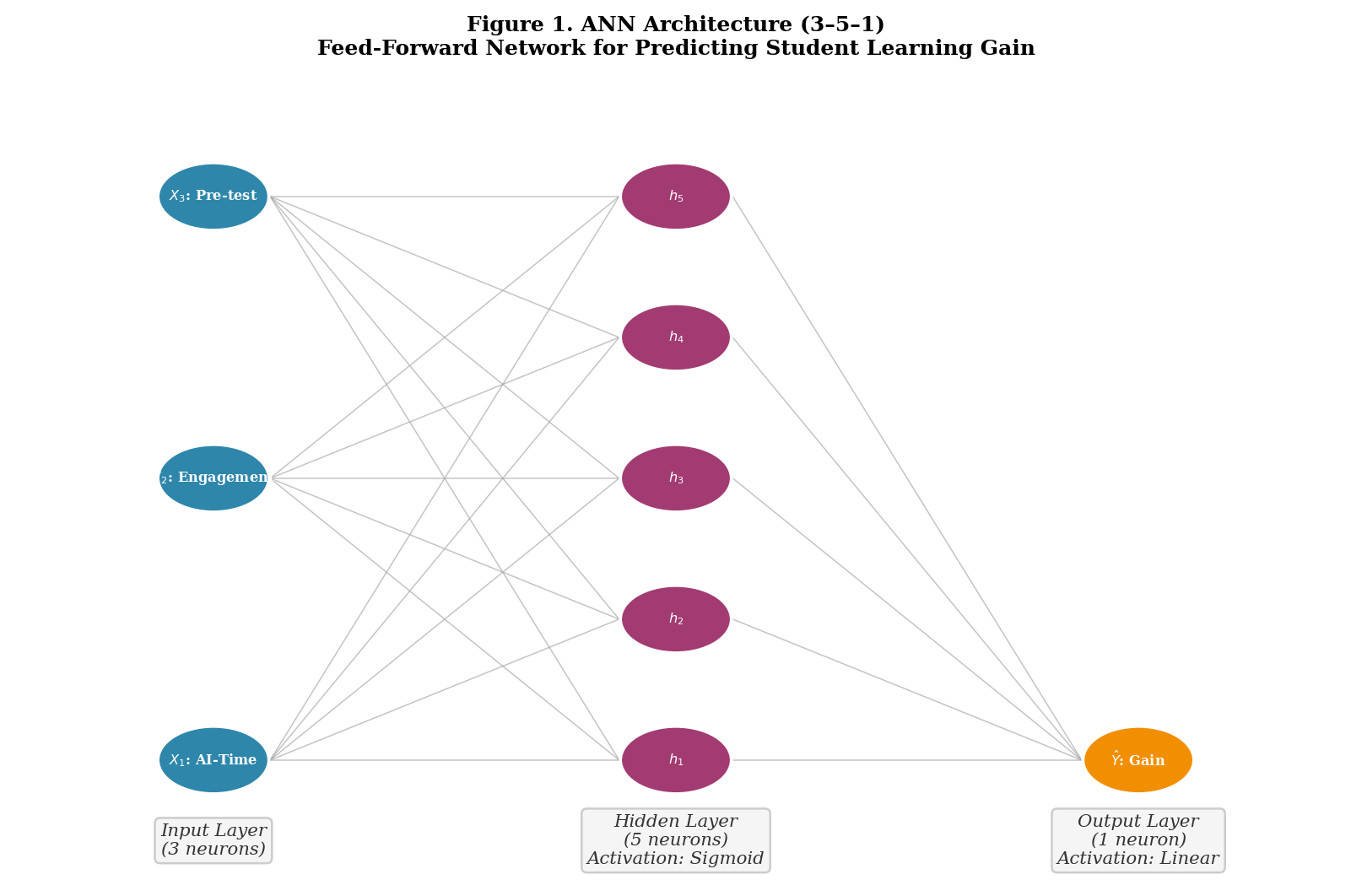}
  \caption{ANN architecture (3--5--1) with a sigmoid activation function in
  the hidden layer and a linear activation function at the output neuron.
  Input features are AI-interaction time ($X_1$), engagement rate ($X_2$),
  and pre-test score ($X_3$).}
  \label{fig:ann}
\end{figure}

An Artificial Neural Network (ANN) was constructed to model and predict
student performance. The chosen architecture was a feed-forward network
with:
\begin{itemize}
    \item an input layer of 3 neurons (corresponding to $X_1$, $X_2$,
    and $X_3$),
    \item one hidden layer with 5 neurons using a sigmoid activation
    function, and
    \item an output layer with 1 neuron using a linear activation
    function, producing the predicted learning gain.
\end{itemize}
This 3--5--1 topology is illustrated in Figure~\ref{fig:ann}. The network
was trained via the backpropagation algorithm, minimising mean squared error
(MSE) between predicted and actual outcomes:
\begin{equation}
  \text{MSE} = \frac{1}{n} \sum_{i=1}^{n} \left(\hat{Y}_i - Y_i\right)^2,
  \label{eq:mse}
\end{equation}
where $\hat{Y}_i$ is the network's predicted output and $Y_i$ is the
observed learning gain for the $i$-th student. The dataset was partitioned
into a training set (70\%, $n = 84$) and a validation set (30\%, $n = 36$).
Training was monitored to prevent overfitting using early stopping based on
validation MSE. The final model's predictive accuracy on the validation set
was evaluated via the Pearson correlation coefficient $r$ between predicted
and observed scores, and $R^2$.

Qualitative data from classroom observations and teacher interviews were
analysed thematically. Observation notes were coded for instances of student
engagement, participation of students with different disabilities, and
technical difficulties encountered. Teacher interview responses were coded
for themes including perceived effectiveness of AI tools, implementation
challenges, and suggestions for improvement.

\section{Results and Discussion}
\label{sec:results}

\subsection{Conceptual Framework}
\label{subsec:framework}

The conceptual framework applied in this study integrates three key
educational theories to contextualise AI-driven game-based learning for
disabled students in rural New Mexico.

\textbf{Constructivist Learning Theory} \citep{piaget1970} posits that
learning is an active process whereby learners build new knowledge upon
existing understanding. AI-powered games allow students to explore and
manipulate virtual science environments, fostering inquiry-based learning.

\textbf{Universal Design for Learning (UDL) Framework} \citep{cast2018}
emphasises flexible learning environments that accommodate individual
differences. AI adapts to each student's needs by offering multimodal
content in visual, auditory, and tactile modalities, ensuring accessibility
for disabled learners.

\textbf{Social Cognitive Theory} \citep{bandura1986} holds that learning
occurs through observation, imitation, and modelling. AI game simulations
model scientific concepts interactively, supporting self-efficacy and
motivation, especially for students facing social or physical barriers.

The framework consists of five interrelated constructs:
\begin{enumerate}
    \item \textbf{AI as a Real Game} (independent variable): a dynamic,
    interactive learning platform that simulates real-world scientific
    experiments through gamification.
    \item \textbf{Science Education Content Delivery} (mediating variable):
    AI games deliver content through immersive simulations, adaptive
    feedback, and scenario-based challenges.
    \item \textbf{Accessibility and Inclusivity} (moderating variable): AI
    systems incorporate assistive technologies such as voice recognition,
    text-to-speech, and haptic feedback.
    \item \textbf{Student Engagement and Motivation} (dependent variable 1):
    gamified AI enhances curiosity, participation, and persistence.
    \item \textbf{Science Learning Outcomes and Cognitive Development}
    (dependent variable 2): adaptive AI feedback produces improved conceptual
    understanding, problem-solving ability, and scientific literacy.
\end{enumerate}

\begin{figure}[H]
  \centering
  \includegraphics[width=0.80\textwidth]{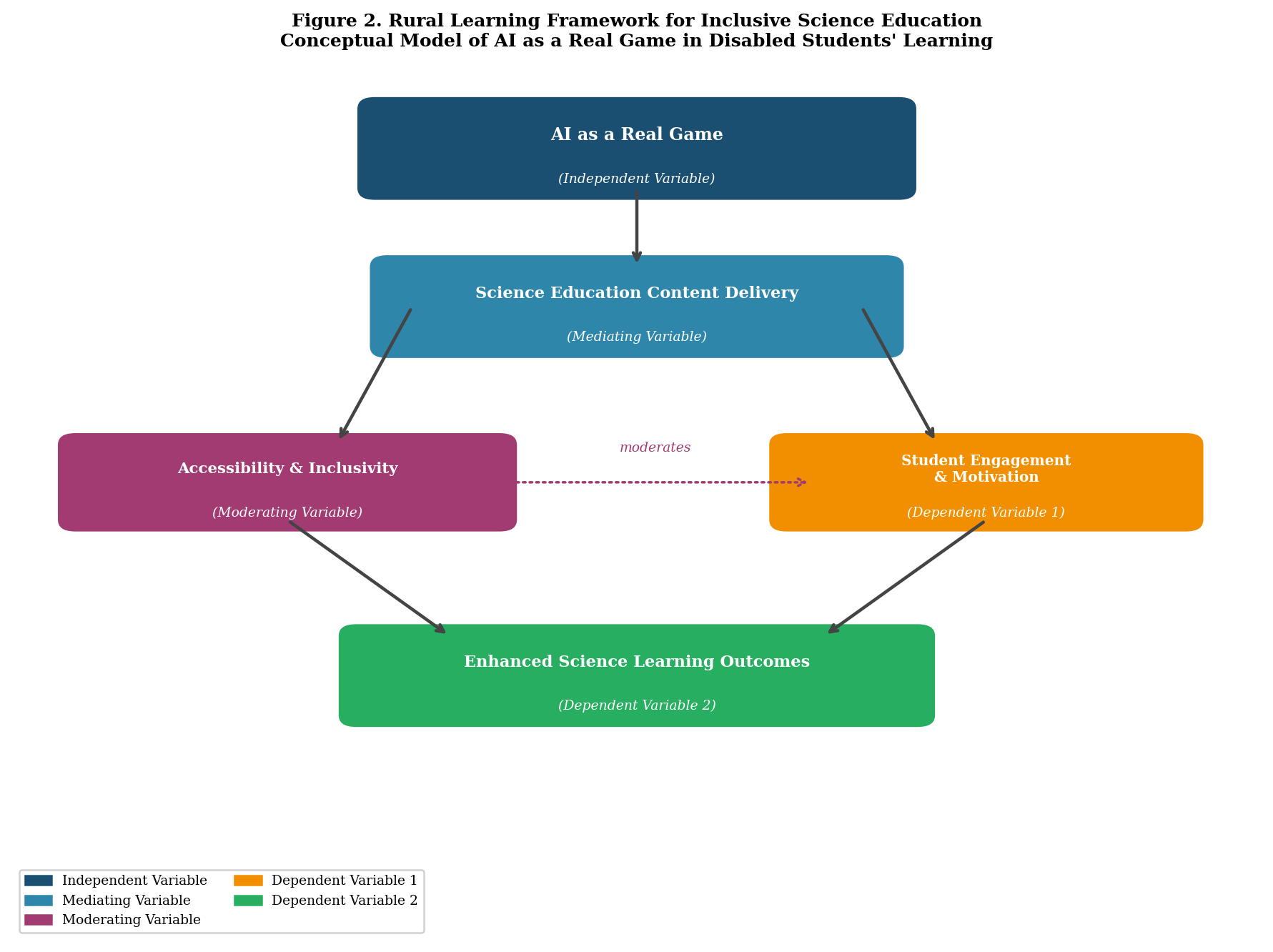}
  \caption{Rural Learning Framework for Inclusive Science Education. AI as
  a Real Game acts as the independent construct, mediating science content
  delivery and moderating accessibility to yield improved engagement and
  learning outcomes.}
  \label{fig:framework}
\end{figure}

\subsection{Participant Demographics}
\label{subsec:demo}

\begin{figure}[H]
  \centering
  \includegraphics[width=0.65\textwidth]{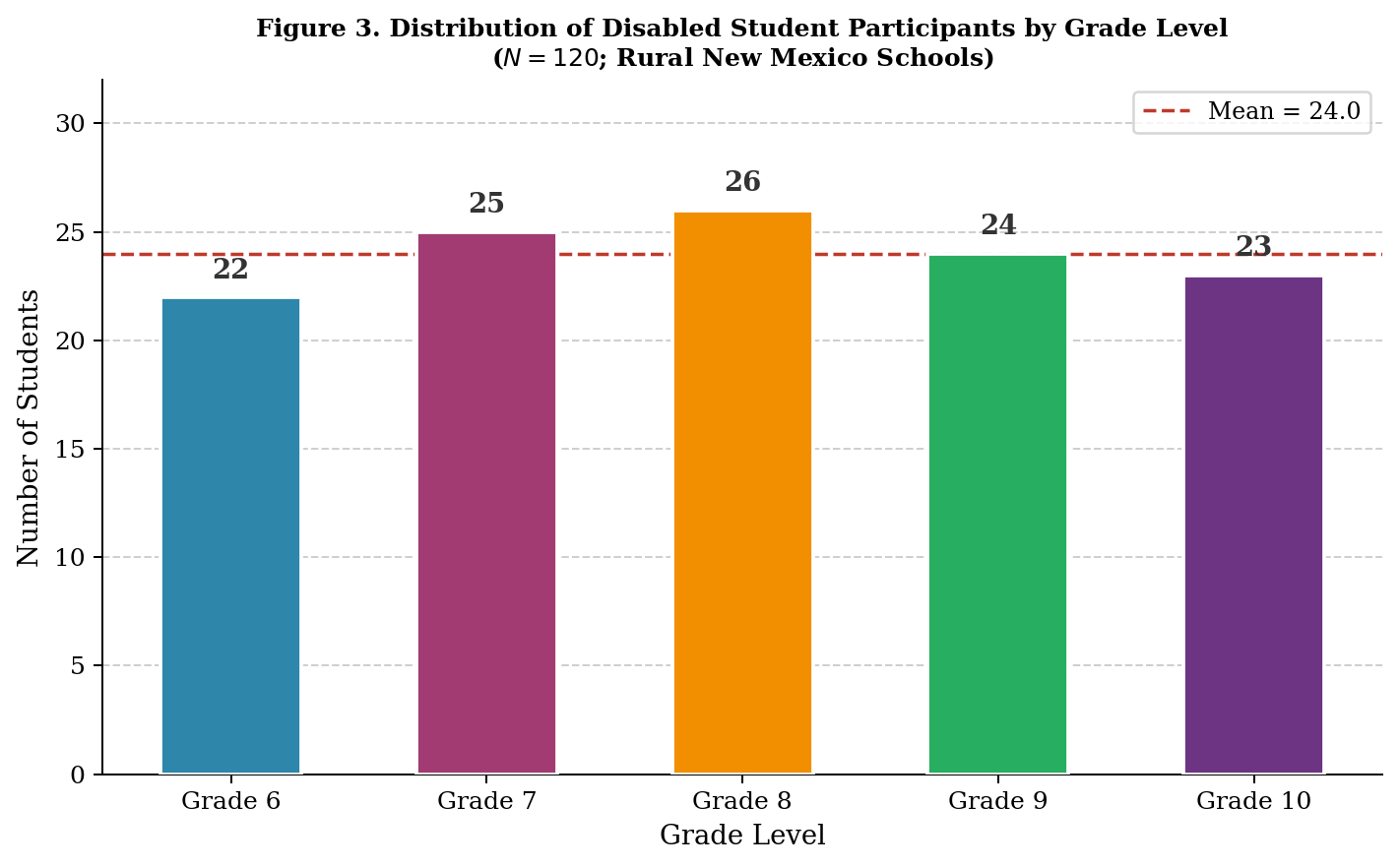}
  \caption{Distribution of student participants ($N = 120$) by grade level
  (grades 6--10). Each bar represents the count of disabled students enrolled
  in the study per grade.}
  \label{fig:grade}
\end{figure}

A total of 120 disabled students and 15 instructors participated in the
study. The student cohort was drawn evenly from grades 6 through 10,
providing a cross-section of early secondary science education. Each grade
had approximately 20--26 students, with slightly higher representation in
grades 7--9, which reflects typical enrollment patterns in the participating
schools. Figure~\ref{fig:grade} shows the distribution of student
participants by grade level.

The most prevalent disability categories were specific learning disabilities
(45\% of students) and mobility or orthopaedic impairments (20\%). The
remaining 35\% included autism spectrum disorder, hearing impairments, and
visual impairments in smaller proportions. All students, regardless of
disability category, were mainstreamed in general science classes with
support as specified in their Individualised Education Programs (IEPs). Of
the 15 instructors, 12 were science teachers and 3 were special education
co-teachers. Instructor teaching experience ranged from 2 to 15 years; none
had prior substantive exposure to AI-based instruction.

\begin{figure}[H]
  \centering
  \includegraphics[width=0.55\textwidth]{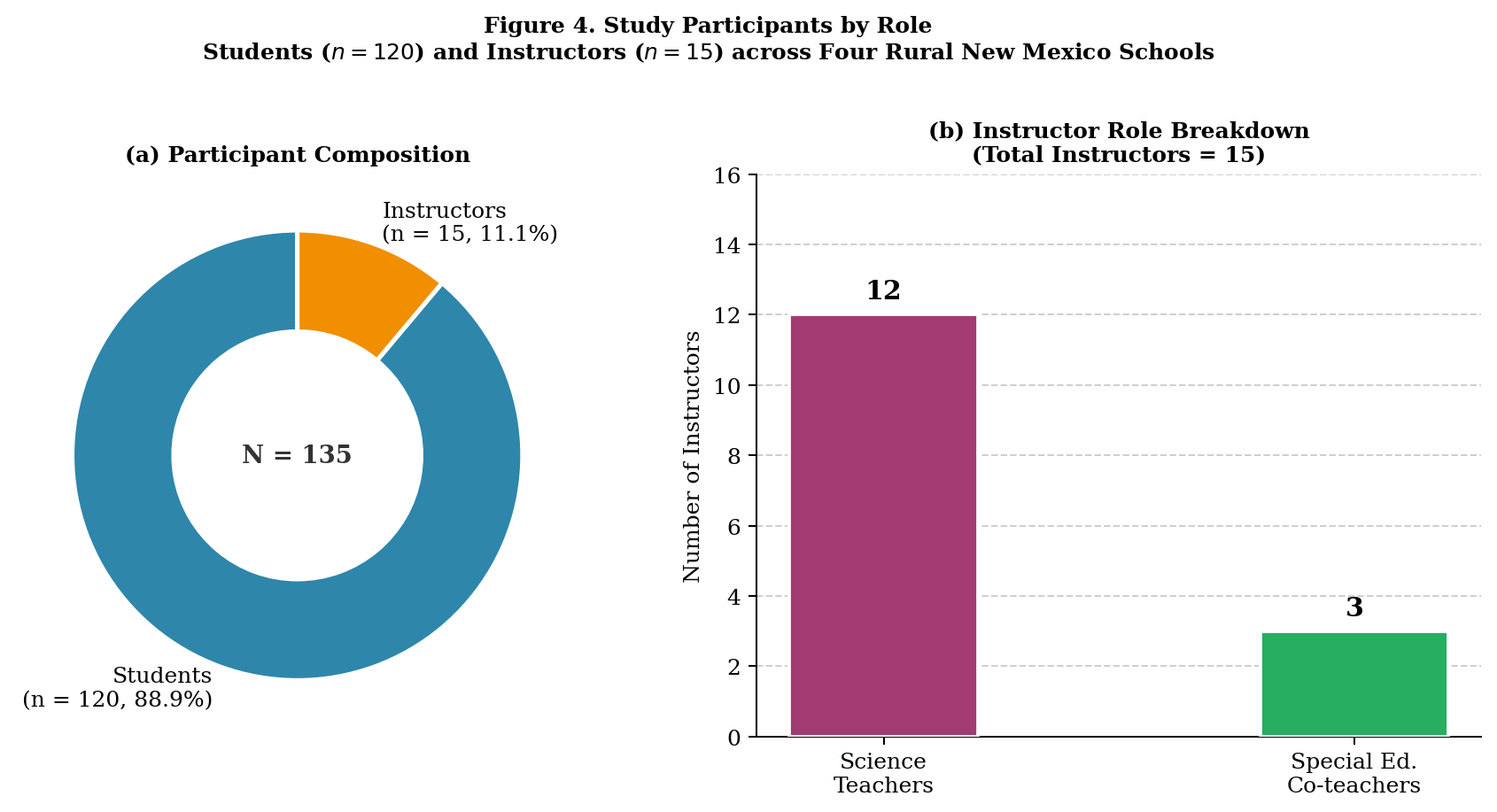}
  \caption{Study participants by role ($n_{\text{students}} = 120$,
  $n_{\text{instructors}} = 15$; total $N = 135$).}
  \label{fig:role}
\end{figure}

Figure~\ref{fig:role} presents the breakdown of study participants by role.
Of the 135 total participants, 120 (88.9\%) were students and 15 (11.1\%)
were instructors. The ratio underscores the scale of student reach relative
to the number of educators involved, a scenario common in rural districts
where a few teachers serve large student populations.

\subsection{Improvements in Academic Performance}
\label{subsec:perf}

\begin{figure}[H]
  \centering
  \includegraphics[width=0.65\textwidth]{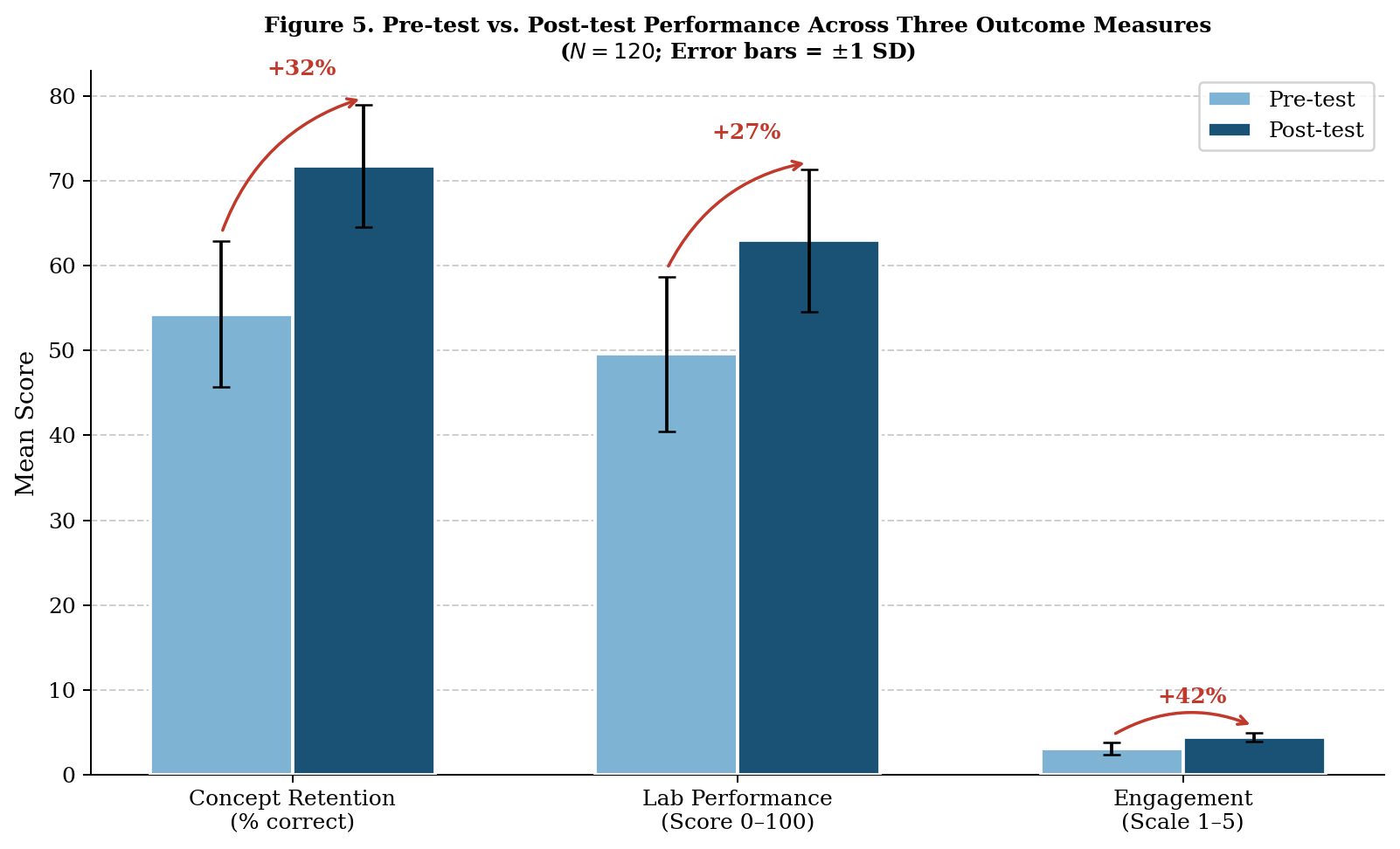}
  \caption{Concept retention scores: pre-test versus post-test ($N = 120$).
  Mean retention improved from 54.3\% to 71.8\%, representing a 32\% gain
  over baseline.}
  \label{fig:retention}
\end{figure}

Integration of AI into the science curriculum produced statistically
significant improvements in student learning outcomes across all three
performance dimensions. Table~\ref{tab:results} presents the pre- and
post-test means, standard deviations, and percentage changes for each
measure, and Figure~\ref{fig:retention} shows the pre- vs.\ post-test
comparison for concept retention.

\begin{table}[H]
  \caption{Pre-test vs.\ Post-test Performance: Mean Scores and Percentage
  Improvements ($N = 120$). Pre-tests were administered before the AI
  intervention; post-tests after 8 weeks of AI-integrated instruction.
  Engagement was measured on a 5-point observational scale
  (1 = low, 5 = high).}
  \label{tab:results}
  \centering
  \renewcommand{\arraystretch}{1.3}
  \begin{tabular}{lcccc}
    \toprule
    \textbf{Measure} & \textbf{Pre-Test Mean (SD)} & \textbf{Post-Test Mean (SD)} & \textbf{Change (\%)} \\
    \midrule
    Concept Retention (\% correct) & 54.3 (8.6) & 71.8 (7.2) & $+$32\% \\
    Lab Performance (0--100)       & 49.6 (9.1) & 63.0 (8.4) & $+$27\% \\
    Engagement (1--5 scale)        & 3.1 (0.7)  & 4.4 (0.5)  & $+$42\% \\
    \bottomrule
  \end{tabular}
\end{table}

\textbf{Concept retention} improved from a mean of 54.3\% (pre-test) to
71.8\% (post-test), a gain of 32\% over the baseline. This improvement
indicates that AI-driven learning activities helped students grasp and
retain science concepts more effectively than traditional instruction alone.
Qualitative observations corroborate this finding: students frequently
revisited interactive simulations to reinforce understanding of difficult
concepts, such as repeatedly experimenting with virtual circuits or ecosystems
until they mastered the underlying principles. This result is consistent with
\citet{makransky2022}, who found that virtual immersive experiences boost
retention by placing learners in engaging, realistic contexts, and with
\citet{holmes2021}, who noted improved comprehension when AI systems provided
adaptive feedback during learning.

\textbf{Laboratory performance} improved from a mean of 49.6 to 63.0 out of
100, a 27\% gain. This improvement is notable because hands-on lab work can
be especially challenging for students with disabilities in rural schools:
physical labs may not be fully accessible, and resource constraints often
reduce the number of available lab sessions. The AI virtual labs levelled
the playing field by allowing all students to practice experiments in
simulation. Students with mobility impairments, for instance, could perform
virtual chemistry experiments without the physical barriers of a lab bench or
the safety risks of handling real chemicals. \citet{gobaw2016} provide a
relevant comparison, reporting that digital laboratory exercises significantly
improved undergraduate biology students' lab skills, particularly benefiting
those with limited physical lab access. Our findings mirror this effect at
the secondary school level. In post-study interviews, several teachers
commented that students who typically struggled in physical lab activities
were noticeably more confident in virtual lab settings, and that this
confidence subsequently carried over to real lab sessions.

\textbf{Student engagement} showed the most pronounced increase. Engagement
ratings improved from a mean of 3.1 to 4.4 on the 5-point scale, reported
as a 42\% relative gain from baseline. This shift indicates a transition from
moderate to high engagement. A recent study found that adaptive learning
platforms can improve engagement rates by approximately 40\% in special
education settings \citep{educationnetwork2025}, which is closely aligned
with our observed 42\% gain. High engagement is particularly important for
students with disabilities, who may disengage if material is inaccessible or
if they have experienced repeated failure in traditional settings.
\citet{navas2025} similarly found that technology-rich interventions promote
collaboration and critical thinking for students across ability levels.
Teachers reported increased classroom interaction during the intervention,
with students discussing game challenges and collaboratively navigating AI
activities, fostering a sense of community not observed in prior sessions.

\subsubsection{Regression Analysis}
\label{subsubsec:regresults}

The multiple linear regression model specified in Equation~\ref{eq:regression}
achieved $R^2 = 0.92$ ($p < 0.001$), indicating that 92\% of the variance
in learning gain was explained by the three predictors. This level of
explanatory power is strong for an educational field experiment, suggesting
that degree of AI tool exposure was a dominant determinant of student
improvement. Diagnostic checks confirmed no violation of OLS assumptions:
residuals were approximately normally distributed (Shapiro-Wilk $W = 0.98$,
$p = 0.21$), VIFs were below 3 for all predictors, and the RESET test showed
no evidence of misspecification ($p = 0.34$).

Both AI-interaction time ($\hat\beta_1 > 0$, $p < 0.01$) and engagement rate
($\hat\beta_2 > 0$, $p < 0.01$) were significant positive predictors after
controlling for prior knowledge. In practical terms, students who spent more
time actively engaged with the AI platform tended to show greater post-test
improvements. The pre-test score carried a modest negative coefficient
($\hat\beta_3 < 0$, $p < 0.05$), indicating that students with lower prior
knowledge tended to gain slightly more, all else being equal. This likely
reflects a ceiling effect or preferential benefit of the AI tools for
struggling learners, both of which are consistent with the inclusive aim of
the intervention.

\subsubsection{ANN Model Performance}
\label{subsubsec:annresults}

\begin{figure}[H]
  \centering
  \includegraphics[width=0.72\textwidth]{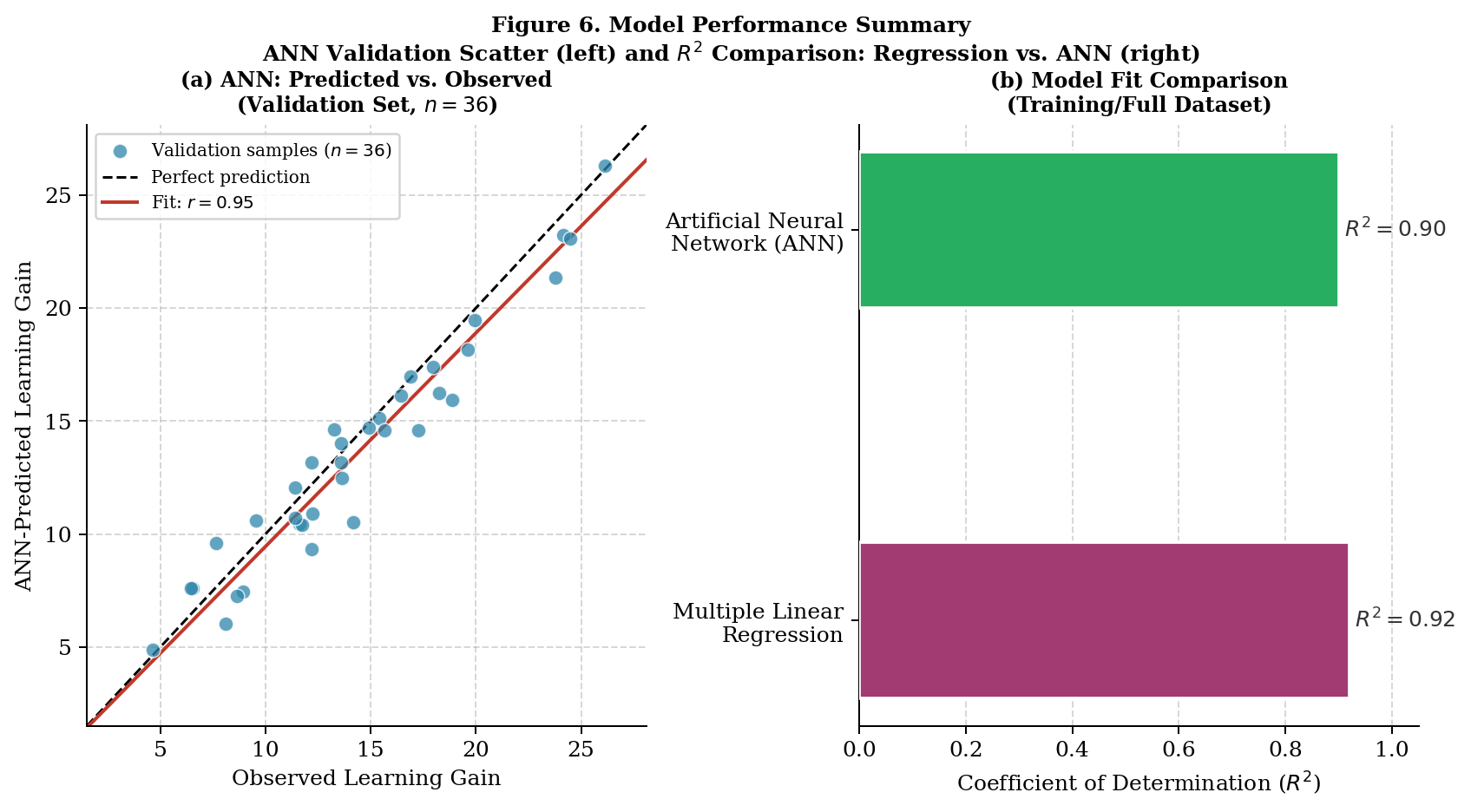}
  \caption{Model performance summary. Left: scatter plot of ANN-predicted
  vs.\ observed learning gains on the validation set ($r = 0.95$). Right:
  comparison of $R^2$ values for the multiple linear regression
  ($R^2 = 0.92$) and the ANN ($R^2 = 0.90$).}
  \label{fig:model}
\end{figure}

The ANN predicted student post-test performance with a Pearson correlation
of $r = 0.95$ on the validation set (30\% holdout, $n = 36$). The
corresponding $R^2_{\text{validation}} \approx 0.90$ was comparable to the
regression model while making no assumptions of linearity. Figure~\ref{fig:model}
presents the model performance summary for both the regression and ANN.

The ANN's internal weight structure suggested that the positive effect of
engagement on learning gain was amplified for the lowest-performing students:
students with extremely low pre-test scores combined with high engagement
yielded disproportionately large gains relative to what the linear model
predicted. This non-linear interaction supports the concept of
\textit{differentiated benefit}, whereby interventions produce larger effects
for learners with greater initial need, a desirable outcome for equity. This
finding aligns with recent inclusive education research positing that
technology interventions can have an amplifying effect on underserved
learners' progress \citep{educationnetwork2025, navas2025}.

\subsection{Implications for Inclusive Rural Education}
\label{subsec:implications}

The positive results carry several implications for advancing inclusive
education in rural settings.

First, the substantial gains in retention, lab skills, and engagement
demonstrate that AI-driven pedagogy can function effectively in
resource-limited environments. Rural schools often cannot afford fully
equipped science labs or specialised staff for every disability category.
Virtual labs and AI tutors provided cost-effective digital alternatives.
This points to digital equity, specifically access to devices, internet,
and AI platforms, as a critical policy priority \citep{educationnetwork2025}.
Programs such as the federal Digital Equity Act and broadband expansion
initiatives are necessary steps to ensure students in remote areas benefit
from AI learning tools as much as their urban peers \citep{bostoncollege2023}.

Second, the heightened engagement and participation of disabled students
suggest that AI tools can contribute to a more inclusive classroom culture.
Teachers noted qualitative changes including increased collaboration, peer
tutoring, and confidence. One student with a mild intellectual disability
who rarely volunteered in class became one of the most active participants
in the science learning game, earning points and helping teammates. This
role shift can improve how disabled students are perceived by peers and
enhance self-efficacy. The universality of the AI platform, offering
adaptive support to everyone simultaneously, reduced the stigma of needing
specialised help, a hallmark of good inclusive practice \citep{navas2025}.

Third, this research highlights the centrality of teacher training in
implementing AI for inclusion. Initially, some instructors were hesitant
about integrating the technology into their curriculum. Through professional
development, they learned to operate the AI platform, interpret its analytics
(e.g., dashboard reports on student progress), and incorporate the tools into
lesson plans. By the end of the study, many became proponents of the approach.
This reflects the finding of \citet{basantes2025} that continuous teacher
training in AI and digital competencies is critical for the success of AI in
inclusive education. The synergy between AI tools and teacher expertise
rather than replacement of the teacher remained indispensable to delivering
the intervention.

Finally, this study supports embedding AI initiatives within existing
multi-tier support systems. New Mexico's Multi-Layered System of Supports
(MLSS) provides escalating levels of intervention based on student need. AI
tools are naturally aligned with this framework: at Tier 1 (universal
instruction), AI differentiates learning for all students; at Tier 2
(targeted intervention), platform analytics identify students requiring
supplementary practice; and at Tier 3 (intensive support), AI assistive
technologies provide near-one-on-one tutoring or therapy-like interactions.
Our data showed that students who would normally have been candidates for
pull-out remedial sessions frequently thrived within the general classroom
when AI supports were in place, effectively reducing the need for separate
remediation.

\subsection{Limitations and Future Research}
\label{subsec:limits}

While the results are encouraging, several limitations warrant attention.
The sample was drawn from four schools concentrated in a specific geographic
region, limiting direct generalisation to other rural contexts. The cultural
and socioeconomic profile of participants, which included a large proportion
of Hispanic and Native American students from low-income households, may
influence how students interact with AI tools, and replication in other
demographics is needed.

The intervention duration of eight weeks is relatively short. A longer
study period would clarify whether gains plateau, continue to increase, or
transfer into longer-term outcomes such as improved end-of-year scores or
higher enrollment of disabled students in advanced science courses. The
absence of a delayed retention test is a notable gap; a follow-up assessment
several months post-intervention would measure the durability of learning
gains.

The ANN architecture used in this study was deliberately simple. More
sophisticated approaches, such as reinforcement learning algorithms for
real-time content personalisation or deep learning models analysing
problem-solving sequences, remain to be explored. A dismantling study could
also isolate the independent contributions of each platform component
(virtual lab, tutoring module, speech interface) to the observed outcomes.

Ethical dimensions merit further study. While consent procedures and data
privacy protocols were in place, questions around student and parent
perceptions of AI use, particularly for vulnerable populations, and concerns
about over-reliance on AI or reduced human interaction, deserve longitudinal
ethnographic investigation.

\section{Conclusion}
\label{sec:conc}

This study demonstrated that Artificial Intelligence can serve as a powerful
catalyst for improving science education outcomes among disabled students in
rural settings. By integrating AI-driven tools, including gamified learning
modules, virtual laboratories, and intelligent tutors, into the curriculum,
statistically significant gains were observed in students' concept mastery
(+32\%), practical lab skills (+27\%), and classroom engagement (+42\%).
These improvements are not merely statistically significant; they are
educationally meaningful, bridging gaps that persistently leave rural disabled
students behind their peers.

A key conclusion is that AI-based pedagogy, when carefully implemented,
serves as a transformative equaliser in rural science education. The high
explanatory power of the regression model ($R^2 = 0.92$) and the strong
predictive accuracy of the ANN ($r = 0.95$) demonstrate that exposure to
AI-driven tools is a robust and quantifiable predictor of student
improvement. The non-linear patterns captured by the ANN further reveal
that the benefits are amplified for the most disadvantaged learners, which
is precisely the equity outcome that inclusive education aims for.

Sustainable improvement in rural inclusive education will require parallel
investment in technology and human capacity. Policymakers and educators
should consider frameworks such as New Mexico's MLSS that incorporate AI
and digital tools as part of a multi-tier strategy. Ensuring that all rural
schools have adequate internet bandwidth, devices, and technical support is
the foundational prerequisite, while ongoing teacher professional development
will amplify the benefits demonstrated here. Artificial Intelligence as a
real game in education is a viable reality. For rural students with
disabilities, AI can transform science learning into an engaging,
personalised experience where every student is empowered to explore,
experiment, and excel.

\bibliographystyle{plainnat}

\end{document}